\documentclass[prb,amsmath,twocolumn,showkeys,showpacs]{revtex4}
\usepackage{graphicx}
\usepackage{dcolumn}
\usepackage{amssymb}
\usepackage{bm}
\begin{document}
\newcommand{\be}{\begin{equation}}
\newcommand{\ee}{\end{equation}}
\newcommand{\caH}{\mathcal{H}}
\newcommand{\la}{\langle}
\newcommand{\ra}{\rangle}
\newcommand{\e}{\epsilon}
\newcommand{\half}{\frac{1}{2}}
\title{ Influences of an impurity on the transport properties of  one-dimensional antisymmetric spin filter}
\author{Jaeshin Park and Hyun C. Lee}
\email[To whom the correspondences should be addressed:~]{ hyunlee@sogang.ac.kr}
\affiliation{Department of Physics and Basic Science Research Institute,
 Sogang University, Seoul, Korea}

\date{\today}
\begin{abstract}
The influences of an impurity on the spin and the charge transport of one-dimensional 
antisymmetric spin filter are investigated using bosonization and Keldysh formulation and the results 
are highlighted against those of spinful Luttinger liquids.
Due to the dependence of the electron spin orientation on wave number the spin transport is 
not affected by the impurity, while the charge transport is essentially identical with that of 
spinless one-dimensional Luttinger liquid. 
\end{abstract}
\maketitle
\section{Introduction}
\begin{figure}[ht]
\begin{center}
\includegraphics[angle=0,width=0.9\linewidth]{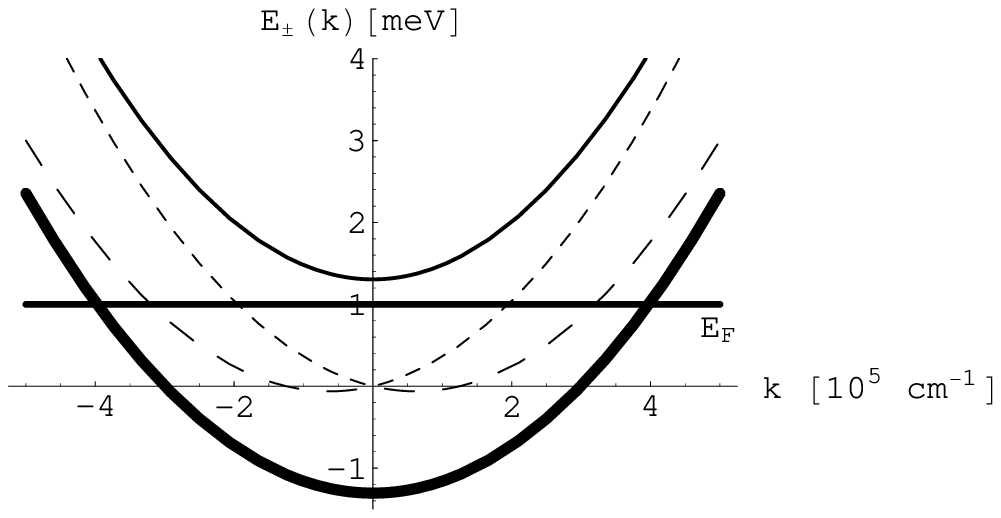}
\includegraphics[angle=0,width=0.9\linewidth]{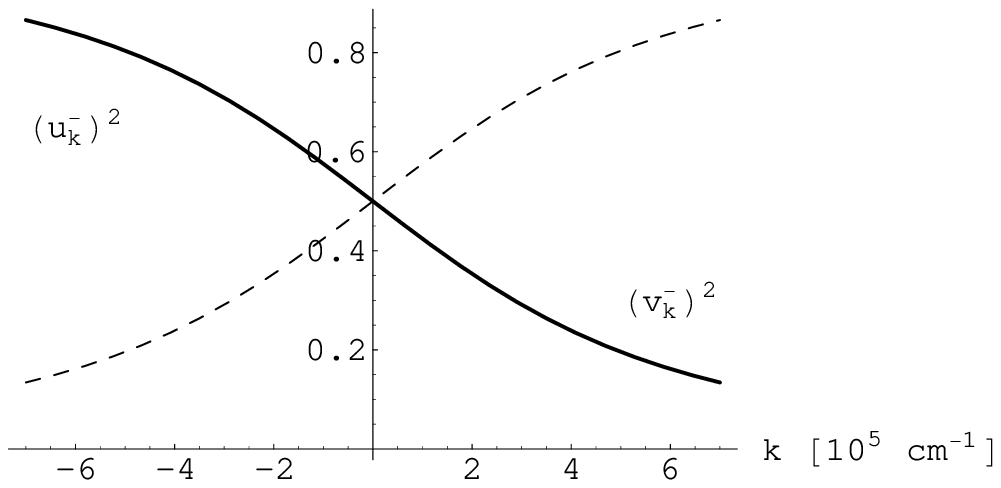}
\caption{
Upper figure: Solid lines represent the lowest energy subband structure of the quantum wire in the absence of the Dresselhaus term.
(Dashed lines are for zero magnetic field).
Note that  the Fermi energy lies in the gap.
In the figure  $B=3 \text{T}$. The $g$-factor is taken to be approximately 15 (as for InAs).
The input parameters are  $\eta_R = 2 \times 10^{-9} \text{eV}\cdot \text{cm}$, $m^*=0.024 m_e$.
Lower figure: The spin-up (solid line) and -down (dashed line) components  
$(u_k^-)^2$ and $(v_k^-)^2$ for the lower $E_-(k)$ band.
The input parameters are identical with the upper figure.
Note that $(v_k^-)^2=1-(u_k^-)^2$. Adapted from Ref.[\onlinecite{mine1}].}
\label{energy}
\end{center}
\end{figure}

The concept of spin filter is an important element in the field of spintronics.\cite{sarma:review,wolf2001}
One of the most representative mechanism of filtering is the spin field effect transistor proposed by
Datta and Das  which is based on the spin-orbit interaction (SOI).\cite{datta1990}
St\v reda and \v Seba \cite{streda2003} proposed an antisymmetric filter (ASF)  which 
employs the Zeeman interaction with \textit{in-plane} magnetic field (or \textit{ parallel} to quantum wire)
as well as Rashba SOI.\cite{rashba1,rashba2}
The interplay of Rashba SOI and the Zeeman interaction with the 
magnetic field parallel to wire gives rise to an interesting one-dimensional (1D)
band structure of quantum wire\cite{streda2003,mine1}, 
where  the orientation of electron spin depends on wave number (see Fig. \ref{energy}).
This dependence on wave number causes the charge and the spin degrees of freedom to mix,
which is a feature distinct from the well-known spin-charge separation of 1D Luttinger liquid (LL).
\cite{bosobook}

The diverse properties of quantum wires in the presence of SOI and/or magnetic field have been
studied: the collective excitations\cite{mine1,collecB}, the interplay of Rashba SOI
and electron-electron interaction\cite{yu2004,gritsev2005}, the optical property\cite{optical}, and
the transmission/reflection coefficients in the presence of a potential step.\cite{imura}
However, as far as we know there is no report on the systematic study of charge/spin transport of 1D
ASF in the presence of impurity scattering and electron-electron interaction.

Impurities necessarily exist in real materials and their effects  are more
pronounced in 1D systems such as quantum wires. Thus, it is important to study the effects of 
impurities in view of the possible realizations of 1D ASF in  low-dimensional nanostructures.
In this paper, we investigate the influences of a single spinless impurity on the 
charge and \textit{spin} transport properties of 1D ASF. 
Remarkably \textit{the spin transport is found not to be affected by the
impurity}, and this is precisely due to the charge-spin mixing effect.
This behavior is in  sharp constrast with that of spinful LL where the spin transport 
is substantially influenced by the impurity scattering [see Eq.(\ref{LL:weak},\ref{LL:strong})].
Contrary to the spin conductance,  the 
charge transport is like that of  \textit{spinless} LL.\cite{kane,akira}
In passing, we mention that in this paper we avoid the delicate problems arising from the contact with leads.

The main results of this paper are the spin and the charge currents of 1D ASF in weak and strong impurity scattering regimes,
which are given by
Eq.(\ref{main1},\ref{main2},\ref{main3},\ref{main4}).

This paper is organized as follows: In Sec. II, we introduce 1D ASF and review the previous results,
in particular the band structure and the bosonized Hamitonian.\cite{streda2003,mine1}
In Sec. III, the impurity Hamiltonian and the coupling to external fields which produce the charge/spin
transport are discussed.
In Sec.IV and V, the bosonized Hamiltonians are analyzed in the framework of Keldysh formalism
and the charge/spin conductances are calculated in the weak scattering and in
the strong scattering regime, respectively. Sec. VI concludes the paper with a summary and discussions.

In this paper we heavily rely on the bosonization method and the Keldysh formulation of transport, and the readers are referred to 
Ref.[\onlinecite{delft}] for the bosonization and Refs.[\onlinecite{weiss,kamenev,rammer}] for the Keldysh method.

\section{One-dimensional antisymmetric spin filter}
\label{asf}
This section is based on Refs.[\onlinecite{streda2003,mine1}], and   the basic setup, the band structure, and the
Hamiltonian of 1D ASF are reviewed.
1D ASF (along x-axis) can be realized 
by applying the confinement potentials  in y and z direction, 
so that the electrons are forced to move along the x-axis.
The confinement in y-direction is due to the Rashba electric field.
We will consider only the lower one-dimensional subband.
Also, a magnetic field is applied along the wire (parallel to x-axis).
The 1D single particle Hamiltonian is given by
\be
\label{one-particle}
\caH_1 = \frac{\hbar^2 k^2}{2 m^*}
+\eta_R k \sigma_z - \epsilon_Z \sigma_x,
\ee
where $\epsilon_Z$ is the Zeeman energy and $\eta_R$ is a parameter characterizing the strength 
of Rashba SOI.
Practically $\eta_R$ is in the range of $(1-10)\times 10^{-9} \mathrm{eV}\cdot\mathrm{cm}$.
By the diagonalization of the Hamiltonian Eq.(\ref{one-particle})
 two bands as depicted in Fig.\ref{energy} are obtained.
When  Fermi energy is located in the gap as shown in Fig.\ref{energy} and 
at low energy, it suffices to take into
account the lower band only.
The energy eigenvalue and
the corresponding normalized eigenspinor of the lower band are given by
\be
\begin{split}
E_-(k) &=\frac{\hbar^2}{2 m^*} k^2 - \sqrt{\e_Z^2 + \eta_R^2 k^2},\cr
\xi_{-} &=\begin{pmatrix} u_k^-  \cr v_k^- \cr \end{pmatrix},
\end{split}
\ee
where  ( $A \equiv \sqrt{(\eta_R k)^2 + \e_Z^2}$ )
\be
\begin{split}
\label{eigenvector2}
u_k^-&=\frac{\e_Z}{\sqrt{(\eta_R k + A)^2 + \e_Z^2}}, \\
v_k^- &=\frac{\eta_R k + A }{\sqrt{(\eta_R k +A)^2 + \e_Z^2}}.
\end{split}
\ee
$u_k^-$ and $v_k^-$ represent the amplitudes for the spin to point in the  +z and the -z direction, respectively.
Fig.\ref{energy} clearly demonstrates that the spin of  left-moving quasiparticles 
is mostly polarized in the +z direction while that of  right-moving quasiparticles is mostly polarized in the -z direction.

Let $a_k$ be the quasiparticle operator of the lower band. At low energy we can neglect 
the quasiparticle excitations of upper band, and the electron operator $c_\sigma$ can be approximately 
expressed in terms of $a_k$ only.\cite{mine1}
\be
\label{simplified}
c^\dag_{k \uparrow}\sim   a_k^\dag u_k^- ,\quad
c^\dag_{k \downarrow} \sim    a_k^\dag v_k^-.
\ee
Also, the $a$-quasiparticle excitations near the left and the right Fermi points 
are more important than others at low energy, so that
the $a$-quasiparticle operator can be decomposed into the 
left ($\psi_L$) and the right moving ($\psi_R$) components.
Then the electron operator $c_\sigma(x)$ can be expressed in terms of $\psi_{R/L}$ 
as follows ($k_F$ is a Fermi momentum):\cite{mine1,note0} 
\begin{align}
\label{operator}
c_\uparrow (x) 
&\sim u_{k_F}^- \, e^{i k_F x} \, \psi_R(x) + u_{-k_F}^- \, e^{-i k_F x} \psi_L(x), \cr
c_\downarrow (x) & \sim v_{k_F}^- \, e^{i k_F x} \, \psi_R(x) + v_{-k_F}^- \, e^{-i k_F x} \psi_L(x).
\end{align}
The non-interacting Hamiltonian in terms of $\psi_{R/L}$ is given by
\be
\label{non-interacting}
\caH_{\mathrm{non}} =v_F \int_{-L/2}^{L/2} dx \Big[ \psi_R^\dag (-i \partial_x) \psi_R +
 \psi_L^\dag (+i \partial_x) \psi_L \Big ],
\ee
where $v_F$ is the Fermi velocity.
The length of 1D ASF is $L$.
The electron-electron interaction Hamiltonian projected on  the lower band is\cite{mine1}
\begin{align}
\label{theHamiltonian}
\caH_{\mathrm{int}} &= 
\frac{g_4}{2}\,\int d x \, \Big[ \rho_R(x) \rho_R(x) +  \rho_L(x) \rho_L(x) \Big ]
\nonumber \\
&+g_2 \int d x \rho_R(x) \rho_L(x),
\end{align}
where $g_4 = V_q$ and $g_2 = V_q - \lambda^2 V_{2 k_F}$.
$V_q$ is a short-range interaction matrix element, so that it is almost momentum-independent.
Here
\be
 \lambda^2 = \frac{\e_Z^2}{\e_Z^2 +( \eta_R k_F)^2},
\ee
and $\rho_{R/L}(x) = \psi^\dag_{R/L}(x) \psi_{R/L}(x)$ is 
the density operator of right/left moving quasiparticles.

The bosonized form of the sum of the non-interacting Hamitonian and the interaction
Hamiltonian is given by\cite{mine1,delft}
\begin{align}
\label{LL:Hamil}
\mathcal{H}_0 &= \pi v_F ( 1+ \frac{g_4}{2\pi v_F})\,\int d x \, \Big[ \rho_R(x) \rho_R(x) +  \rho_L(x) \rho_L(x) \Big ] \nonumber \\
&+g_2 \int d x \rho_R(x) \rho_L(x).
\end{align} 
It is convenient to define the LL parameter $K$ and the velocity of collective excitation
$v_0$.
\begin{align}
\label{parameters}
K & = \sqrt{ \frac{1+ \frac{g_4}{2\pi v_F} - \frac{g_2}{2\pi v_F}}{1+ \frac{g_4}{2\pi v_F} + \frac{g_2}{2\pi v_F} }}, \cr
v_0 &= v_F \sqrt{ ( 1+ \frac{g_4}{2\pi})^2 - ( \frac{g_2}{2\pi})^2}.
\end{align}
For a repulsive electron-electron interaction, $K < 1$.
The action corresponding to the Hamiltonian Eq.(\ref{LL:Hamil}), being expressed in terms of phase 
fields, is given by
\be
\label{theaction}
S_0 = -\int dt \, dx \Big[ \frac{1}{\pi} \partial_x \phi \partial_t \theta +
\frac{v_0}{2\pi}\,\Big( K (\partial_x \phi)^2 + \frac{1}{K}\,(\partial_x \theta)^2 \Big ) \Big ],
\ee
where the phase fields $\theta$ and $\phi$ are defined by the following relations\cite{delft}
\begin{align}
\rho_R + \rho_L  &= \frac{1}{\pi} \partial_x \theta + \frac{\widehat{N}_R +\widehat{N}_L}{L} , \cr
\rho_R -\rho_L  &=\frac{1}{\pi} \partial_x \phi + \frac{\widehat{N}_R -\widehat{N}_L}{L}.
\end{align}
$\widehat{N}_{R/L}$ is the total number operator of right/left moving fermions.
\section{Impurity Hamiltonian and coupling with external fields}
The scattering by a spinless impurity (located at $x=0$) is described by the following Hamiltonian.
\be
\label{impurity1}
\caH_{\mathrm{imp}} = V_0  \sum_{\sigma = \uparrow, \downarrow}\, c_\sigma^\dag(x=0) c_\sigma(x=0).
\ee
Projected on the lower band using Eq.(\ref{operator}) 
the impurity Hamiltonian Eq.(\ref{impurity1}) becomes 
\be
\label{impurity2}
\caH_{\mathrm{imp}} = (V_0 \lambda) (2\pi a)\,\big[ \psi_R^\dag(0) \psi_L(0) 
+  \psi_L^\dag(0) \psi_R(0)\big ],
\ee
where $a$ is a short distance cutoff of the order of lattice spacing and 
the unimportant forward scattering terms are omitted.
Note that the backscattering amplitude is suppressed by a factor $\lambda$ which is just a 
overlap of two spinors at $k=\pm k_F$. Thus, this suppression is  a consequence of charge-spin
mixing.

Employing the bosonization formula we get
\be
\label{impurity3}
\caH_{\mathrm{imp}} = W_0 \Big( F_R^\dag F_L e^{-i 2 \theta(0,t)} + \mathrm{H.c} \Big ),
\ee
where $W_0 = ( \lambda V_0) (2\pi a)$ and $F_{R/L}$ is the Klein factor\cite{delft}.
The renormalization group flow of impurity scattering strength $W_0$ with 
the Hamiltonian Eq.(\ref{LL:Hamil}) and Eq.(\ref{impurity3}) is  well understood.\cite{kane,akira}
The scaling equation is ($dl  = - \frac{d \Lambda}{\Lambda}$)
\be
\label{scaling}
\frac{d W_0(l)}{d l} = (1 - K ) W_0(l),
\ee
where $\Lambda$ is the flowing  energy cutoff of the system.
If $K < 1$ (the repulsive electron-electron interaction), the impurity scattering 
becomes stronger at lower energy. Therefore, it is natural to divide the problem into two regimes:
the weak scattering (or high temperature) regime where the impurity scattering can be treated perturbatively and 
the strong scattering (or low temperature) regime where we had better start from 
two disconnected quantum wires which are weakly linked
by tunnelings at finite temperature.\cite{kane,gogolin}

We will compute the charge and the spin transport in two regimes. For transport to occur,
some external fields should be applied.
For the charge transport we will apply the \textit{potential difference} \cite{akira}
($V(x) = -\frac{V}{2} \mathrm{sign}(x)$) across the impurity.
Similarly, for the spin transport, the \textit{magnetic field difference} along z-axis \cite{akira}
( $\vec{B}_{\mathrm{p}}(x) = \frac{B_0}{2} \mathrm{sign}(x) \widehat{z}$ ) is applied
across the impurity.
We emphasize that $\vec{B}_{\mathrm{p}}(x)$ has to be distinguished from the magnetic field applied 
parallel to the wire (along x-axis) which is necessary for the construction of ASF itself.
The probe magnetic field can be applied in arbitrary direction in y-z plane, in general.
It turns out that the contribution  coming from the y-component of $\vec{B}_p$
is multiplied by  
a oscillating factor $e^{ \pm i 2 k_F x}$, so that it becomes negligible upon spatial integration.

The Hamiltonian for the interaction with the potential difference is
\be
\caH_{V} =  \sum_{\sigma=\uparrow,\downarrow} \int dx\, (-e) V(x)  c^\dag_\sigma(x) c_\sigma(x).
\ee
After the projection on the lower band using Eq.(\ref{operator}), 
the bosonized form of  $\caH_{V}$ is given by\cite{boundaryterm}
\be
\label{potential}
\caH_{V} = \frac{e V}{\pi} \theta(x=0,t).
\ee

The Hamiltonian for the interaction with the  magnetic field difference in the z-direction is
($\uparrow = +1, \downarrow = -1$)
$$\caH_B = \sum_{\sigma=\pm 1} \, \int dx\,\mu_B B_{\mathrm{p}}(x) \sigma \,c^\dag_\sigma(x) c_\sigma(x).$$
Again after the projection to the lower band using Eq.(\ref{operator})  we obtain
\begin{align}
\label{magnetic3}
&\caH_B =\int dx\,\mu_B B_{\mathrm{p}}(x)\,\Big[ [(u_{k_F}^-)^2 - (v_{k_F}^-)^2] \psi^\dag_R(x) \psi_R(x) \cr
&+[(u_{-k_F}^-)^2 - (v_{-k_F}^-)^2] \psi^\dag_L(x) \psi_L(x) \Big ].
\end{align}
A short computation shows that
\be
(u_{k}^-)^2 - (v_{k}^-)^2 = - \frac{ \eta_R k}{\sqrt{\epsilon_Z^2 + (\eta_R k)^2}}.
\ee
Thus we arrive at ($\kappa = \frac{ \eta_R k_F}{\sqrt{\epsilon_Z^2 + (\eta_R k_F)^2}}$ )
\be
\label{magnetic}
\begin{split}
\caH_B &= - \kappa \mu_B\, \int dx  B_p(x) \Big[ \psi_R^\dag(x) \psi_R(x) - \psi_L^\dag(x) \psi_L(x)\Big ] \cr
&= -\kappa \mu_B \int dx B_p(x) \frac{1}{\pi} \partial_x \phi(x).
\end{split}
\ee
Evaluating the integral of Eq.(\ref{magnetic}) we obtain
\be
\label{magnetic2}
\caH_B = \frac{ \kappa \mu_B B_0}{\pi} \phi(x=0,t).
\ee
In the case of 1D ASF the couplings to the electric potential and the magnetic field are \textit{not} independent from
each other because $[\theta,\phi] \neq 0$ in general. This is a feature which is very different from
that of  spinful LL with spin-charge separation. Elaborating on this point further,
it is interesting to compare the Hamiltonian Eq.(\ref{potential}, \ref{magnetic2}) with those of
spinful LL:\cite{akira,note2}
\be
\caH_{V,sLL} = \frac{e V \theta_\rho(0)}{\pi}, \quad
\caH_{B,sLL} = \frac{ \mu_B B_p \theta_\sigma(0)}{2\pi},
\ee
where $\theta_\rho$ and $\theta_\sigma$ are the charge/spin boson phase field
which describes the fluctuations of 
charge/spin density, and they are independent from each other in the sense of $[\theta_\rho,\theta_\sigma]=0$.

To find the charge and the spin current we note that the electric potential couples to the charge and 
the magnetic field couples to the magnetic moment. 
Then from Eq.(\ref{potential}) and Eq.(\ref{magnetic2}), 
the following expressions of charge/spin currents can deduced.
\be
\label{chargecurrent}
J_\rho = \frac{(-e)}{\pi}\, \frac{ d \la  \theta(0,t) \ra}{d t}.
\ee
\be
\label{spincurrent}
J_\sigma = (-1) \frac{\kappa}{\pi} \frac{d \la \phi(0,t) \ra}{d t}.
\ee
Here  $\la  \theta(0,t) \ra$ and $ \la \phi(0,t) \ra$ are the averages over the non-equilibrium ensemble.
The Keldysh formalism will be employed in computing these non-equilibrium averages.

\section{Weak scattering regime} 
\label{weak}
The total Hamiltonian of the system is [see Eqs.(\ref{LL:Hamil},\ref{impurity3})]
\be
\caH = \caH_0 + \caH_{\mathrm{imp}} + \caH_S,
\ee
where $\caH_S$ is the source Hamiltonian for the coupling to  external field. 
For the computation of charge transport
$\caH_S = \caH_V$ [Eq.(\ref{potential})], and for the computation of spin transport $\caH_S = \caH_B$  
[Eq.(\ref{magnetic2})].
In the weak scattering regime we can treat $\caH_{\mathrm{imp}}$ perturbatively.
In Keldysh path integral formulation the key element is the following functional integral\cite{weiss,kamenev,rammer}
\be
Z = \int D[\theta_{f,b}, \phi_{f,b}] \,e^{i S_K + i S_{\mathrm{imp}} + i S_S},
\ee
where $(\theta,\phi)_{f,b}$ denote the phase boson fields defined on the forward time branch and the backward
time branch of the closed time contour, respectively.
$S_K$ is the Keldysh action corresponding to the Hamiltonian Eq.(\ref{LL:Hamil}).
It is basically the difference of the action Eq.(\ref{theaction}), $S_K = S_{0,f}-S_{0,b}$ between the forward and the backward branch, which are eventually expressed in terms of $\theta_{c/q}=(\theta_f\pm \theta_b)/2$ and $\phi_{c/q}=(\phi_f \pm \phi_b)/2$.
 $S_{\mathrm{imp}}$ and $S_S$ is the Keldysh action for
 impurity Hamiltonian and the source Hamiltonian $\caH_S$, respectively.

\underline{The spin transport} - The spin current Eq.(\ref{spincurrent}) can be computed easily
by the coupling to external sources.\cite{kamenev}
The Hamiltonian Eq.(\ref{magnetic2}) expressed in the form of Keldysh source 
action is
\be
S_B = - \frac{\kappa \mu_B}{\pi} \int_{-\infty}^\infty dt \,\Big [
 B_{0q}(t) \phi_{c}(0,t) + B_{0c}(t) \phi_{q}(0,t) \Big ],
\ee
where $B_{0 c/q}$ is the classical/quantum component of external magnetic field.\cite{kamenev}
The spin current in the 0-th order of impurity scattering is 
\be
J_\sigma^{(0)}(t) = i \mu_B^{-1} \frac{d}{d t}\left( \frac{ \delta Z^{(0)}[B_{0c},B_{0q}]}{\delta B_{0q}(t)} \Bigg \vert_{B_{0q} =0} \right ),
\ee
where $Z^{(0)}[B_{0c},B_{0q}] = \la e^{i S_B} \ra$. Here the average means the Keldysh functional integral 
with respect to $S_K$. Since $S_K$ is Gaussian in $\theta_{c,q}$ and $\phi_{c,q}$ we can use the identity
$\la e^{ i X } \ra = e^{-\la X X \ra/2}$. Employing an identity $\la \phi_q \phi_q \ra =0$, we find
\begin{align}
&J_\sigma^{(0)}(t)  = \frac{i}{\mu_B} ( \frac{\kappa \mu_B}{\pi})^2 
\frac{d }{d t} \Big[ \int_{-\infty}^\infty dt' \cr
\times &( \la \phi_q(0,t') \phi_c(0,t) \ra B_{0c}(t') +
\la \phi_c(0,t) \phi_q(0,t') \ra B_{0c}(t') ) \Big ] \cr
&=\mu_B \frac{\kappa^2}{2\pi K} B_0,
\end{align}
where we have used ($\Theta(t)$ is a Heaviside step function)
\begin{align}
\label{phi-function}
\la \phi_c (0,t_1) \phi_q (0, t_2) \ra &= - \frac{i \pi}{4 K} \Theta(t_1 - t_2), \cr
\la \phi_q (0,t_1) \phi_c (0, t_2) \ra &= - \frac{i \pi}{4 K} \Theta(t_2 - t_1). 
\end{align}
The first order correction to the spin current by impurity scattering  vanish  
because a single Klein factor does not conserve fermion number.
As of the second order correction, the above argument does not work since
$F_{L/R} F^\dag_{L/R}=1$ conserves the fermion number.
The second order correction is schematically given by
\be
J_\sigma^{(2)} \propto \frac{d}{dt}\,\left( \frac{\delta}{\delta B_q(t)}\,
\la e^{i S_B} S_{\mathrm{imp}}  S_{\mathrm{imp}} \ra \right )
\ee
In view of the fact that the impurity scattering is proprotional to $e^{ 2 i \theta(0,t)}$ we find
that the functional differentiation would generate the  Green functions only of the type 
$\la \theta_{c/q}(x = 0,t) \phi_{c/q}(x=0,t') \ra$ which vanishes identically.
 Note that this vanishing of the second order correction is \textit{solely due 
to the specific form of the Hamiltonian Eq.(\ref{magnetic2}) whose origin can be traced back to
the spin-charge mixing effect of 1D ASF}. Summarizing the result,
\be
\label{main1}
J_\sigma = \mu_B \frac{\kappa^2}{2\pi K} B_0 = J_\sigma^{(0)},\;\;J_\sigma^{(1)} = J_\sigma^{(2)}=0.
\ee
This is one of the main results of this paper. 
The corrections can only stem from the failure of linearization approximation which is necessary for the 
bosonization approach, therefore, such corrections
are expected to be very small at low temperature.
From Eq.(\ref{main1}) the spin conductance easily follows.
\be
\label{main1-conductance}
G_\sigma \equiv \lim_{B_0 \to 0} \frac{J_\sigma}{B_0} = \mu_B \frac{\kappa^2}{2\pi K}, \;\;
\text{no corrections}.
\ee

\underline{The charge transport} - The source Hamiltonian necessary for the computation of
the charge current is given by Eq.(\ref{potential}) which does \textit{not} depend on the field $\phi$.  
With $\phi$  integrated out, the
action Eq.(\ref{theaction}) becomes
\be
\label{theta-action}
S_\theta =\frac{1}{2\pi K} \int dt dx \Big( \frac{1}{v} (\partial_t \theta)^2 -
v (\partial_x \theta)^2 \Big).
\ee
This is the action for the \textit{spinless} LL with LL parameter $K$. 
The charge transport based on the action Eq.(\ref{theta-action}) have been calculated 
by the linear response theory\cite{kane} and by the influence functional method\cite{akira}.

The calculation of the 0-th order charge current is entirely identical with that of the 
spin current except for the subsitution of $\phi \to \theta$ and the change of  parameters.
\be
\label{chargezero}
J_\rho^{(0)} = \frac{e^2 K }{ 2\pi } V.
\ee
This is just the charge conductance of \textit{one-channel (or spinless)} quantum wire.
The first order correction due to impurity scattering vanishes again due to Klein factor.
The second order correction  is given by
\begin{align}
\label{secondorder}
&J_\rho^{(2)} \propto i \frac{d}{d t} \Bigg[ \frac{\delta}{\delta V_q(t)} \int dt_1 dt_2 
\Big \la (e^{2i \theta_f(t_1) } - e^{2i \theta_b(t_1) }) \cr
& \times(e^{-2i \theta_f(t_2) } - e^{2i \theta_b(t_2 ) }) 
e^{ \frac{e}{\pi}\,i \int d t' (V_q \theta_c +V_c \theta_q)} \Big \ra \Big \vert_{V_q =0}\Bigg ],
\end{align}
where the average is done with respect to the Keldysh action $S_K$. The average is  Gaussian
functional integration, and 
 the result turns out to be
\begin{align}
\label{chargesecond}
J_\rho^{(2)} &\sim W_0^2 \int_{-\infty}^\infty dt' e^{-2 K C(t') }\,
\sin ( \frac{e V K t'}{2}) \sin ( \frac{ \pi K}{2} \mathrm{sign}(t')) \cr
 &\sim W_0^2  \int_{0}^\infty dt'\frac{ \sin (\frac{e V t'}{2})}{
(t^{'2}+\tau_c^2)^{K} ( \frac{ \sinh ( t' \pi /\beta)}{t' \pi /\beta} )^{2K}},
\end{align}
where  
\begin{align}
\label{C-function}
C(t') &\equiv \int_0^\infty \frac{d \omega}{\omega}\,e^{-\omega \tau_c}
 \coth\frac{\beta \omega}{2} [1-  \cos \omega (t')] \cr
&= \frac{ \sqrt{(t')^2 +\tau_c^2}}{\tau_c} +  \ln \left [   \frac{ \sinh \frac{  t' \pi }{\beta}}{
\frac{  t' \pi }{\beta}}
\right ],
\end{align}
where $\tau_c$ is a short-time cutoff. Collecting the previous results, we get
(up to the second order in $W_0$ )
\be
\label{main2}
J_\rho = \frac{e^2 K}{2 \pi} \Big[ V - c_\rho W_0^2  \int_{0}^\infty dt'\frac{ \sin (\frac{e V K t'}{2})}{
(t^2+\tau_c^2)^{K} ( \frac{ \sinh ( t \pi /\beta)}{t \pi /\beta} )^{2K}} \Big ].
\ee
$c_\rho$ is a constant.
Note that this expression is essentially identical with that by Fisher and Zwerger (
Eq.(3.51) of Ref.[\onlinecite{zwerger}] ).
From Eq.(\ref{main2}) the charge conductance easily follows:
\be
\label{main2:conductance}
G_\rho = \lim_{V \to 0} \frac{J_\rho}{V} = \frac{ e^2 K }{2\pi} \Big[1 - \tilde{c}_\rho T^{2K-2} \Big ].
\ee
It is very interesting to highlight our results on ASF against those of spinful LL.
From Eq.(3.15) and Eq.(3.18) of Ref.[\onlinecite{akira}] we have
\begin{align}
\label{LL:weak}
G_{\rho,sLL} & = \frac{ e^2 K_\rho}{\pi} \Big[ 1 - c_1( \frac{\pi T}{\Lambda})^{K_\rho+K_\sigma - 2} + \cdots \Big ], \cr
G_{\sigma,sLL} & = \frac{ \mu_B K_\sigma}{\pi}
 \Big[ 1 - c_2 ( \frac{\pi T}{\Lambda})^{K_\rho+K_\sigma - 2} + \cdots \Big ],
\end{align}
where only the leading terms are indicated. $K_\rho$ and $K_\sigma$ is the LL parameter for the 
charge and the spin degrees of freedom, respectively. $c_{1,2}$ are constants.

The comparison of the charge conductance Eq.(\ref{main2:conductance}) 
with Eq.(\ref{LL:weak}) shows that the charge transport of 1D ASF essentially behaves like that of 
\textit{spinless LL}. However, the LL parameter $K$ depends sensitively on the Rashba SOI and the 
Zeeman interaction (recall that $g_2$ depends on them).
The spin conductance of 1D ASF Eq.(\ref{main1-conductance}) is \textit{qualitatively} different from that of spinful LL Eq.(\ref{LL:weak}). 
The absence of corrections to the spin conductance of 1D ASF reflects the dependence of spin orientation on 
wave number. The backscattering reverses the momentum, and this degrades charge  flow. However,
from the viewpoint of spin, the momentum reversed state has the spin orientation which is almost parallel
to the one in the absence of impurity, so that the spin current does not degrade. 
Even the electron-electron interaction can not modify this property significantly.

\section{Strong scattering regime}
\label{strong}
As mentioned in Sec. III, the proper starting point in the strong scattering regime at zero temperature
is two disconnected semi-infinite wires. Finite temperature and external fields make
tunneling between two wires possible, and it results in transport.
The 1D interacting system with boundary is  most conveniently described 
by the open-boundary bosonization.\cite{gogolin}
Let us designate two disconnected wires by 1 and 2.
For each semi-infinite wire, the boundary condition at the end ($x=0$) relates the left and right moving 
electrons, so that the left moving fields can be expressed solely in terms of right moving fields
(as  reflected images)
\be
\psi_{a L}(x) = - \psi_{a R}(-x), \quad \rho_{a L}(x) = \rho_{a R}(-x),\;a =1,2.
\ee

The bosonized Hamiltonian of each wire which is expressed purely in terms of the right moving fields are 
\begin{align}
\label{open:hamil}
\mathcal{H}_a &= \pi (v_F+ \frac{g_4}{2\pi} )\, \int_{-L/2}^{L/2} dx \rho_{a R}^2(x)  
\cr
&+ \frac{g_2}{2} \, \int_{-L/2}^{L/2} dx \rho_{a R}(x)  \rho_{a R}(-x),\;\; a = 1, 2.
\end{align}
Note that the last term of Eq.(\ref{open:hamil}) is \textit{non-local in space}.
It is interesting to compare Eq.(\ref{open:hamil}) with Eq.(\ref{LL:Hamil}) and to notice how 
the presence of boundary is reflected in the structure of the Hamiltonian.
The density operator in terms of \textit{chiral} boson field is given by
\be
\label{density}
\rho_{a R}(x) = \frac{\widehat{N}_a}{L} + \frac{1}{2\pi}\, \partial_x \phi_{a R}(x),
\ee
where $\widehat{N}_a$ is the fermion number operator of the $a$-th wire.
The tunneling between two wires is given by \cite{kane,gogolin}
\be
\mathcal{H}_T = t_u \Big [ F_{1 R} F_{2 R}^\dag e^{i \phi_{1R}(x=0) - i \phi_{2 R}(x=0)} + 
\mathrm{H. c} \Big ].
\ee

The coupling to the potential difference 
is described by the Hamiltonain
\be
\label{potential-s}
\mathcal{H}_{V,s} =  \frac{e V}{2} ( \widehat{N}_1 - \widehat{N}_2).
\ee
The comparison of two Hamiltonians Eq.(\ref{potential}) and Eq.(\ref{potential-s}) reveals an important difference 
in the charge transport mechanism between the weak and the strong scattering regime.
The Hamiltonian in the weak scattering regime Eq.(\ref{potential}) is given in terms of \textit{phase field} $\theta$ which commutes with the Klein factors, while the Hamiltonian in the strong scattering regime Eq.(\ref{potential-s})
does \textit{not} commute with the Klein factors. Because of this property it is not feasible to apply the 
Keldysh formalism on the charge transport in the strong scattering regime.

As for the coupling with the magnetic field difference, starting from Eq.(\ref{magnetic3}) one can derive
\begin{align}
\mathcal{H}_{B,s} &= - \kappa \mu_B\,\Big[ \int_{-L/2}^0 dx \frac{-B_0}{2} [
\rho_{1R}(x)  - \rho_{1 L}(x) ]  \cr
&+\int^{L/2}_0 dx \frac{B_0}{2} [
\rho_{2 R}(x)  - \rho_{2 L}(x) ]  \Big ] 
\end{align}
Using  Eq. (\ref{density}) (we set $\phi_{a R}(x=\pm L/2) =0$) we obtain
\be
\mathcal{H}_{B,s} = \frac{\kappa \mu_B B}{2 \pi}\, \Big[ \phi_{1R}(x=0) + \phi_{2R}(x=0) \Big ].
\ee
The examination of the tunneling and external field Hamiltonians necessitates the introduction of the symmetric and the antisymmetric combinations of operators.
\be
\phi_\pm = \frac{\phi_{1R} \pm \phi_{2R}}{\sqrt{2}}, \;\;
\widehat{N}_\pm = \frac{ \widehat{N}_1 \pm \widehat{N}_2}{2}.
\ee
 Let us also define $F \equiv F_{1R} F_{2R}^\dag$ which satisfies the following relations.
\be
F F^\dag = F^\dag F = 1,\;\; [N_-, F] = - F, \;\; [N_+, F] = 0.
\ee
In terms of these new fields,
\begin{align}
\label{allofthem}
\mathcal{H}_T &= t_u \Big[ F e^{ i \sqrt{2} \phi_-(x=0)} + \mathrm{H.c} \Big ], \cr
\mathcal{H}_{V,s} &= - e V \widehat{N}_-, \quad
\mathcal{H}_{B,s}  = \frac{\kappa \mu_B B_0}{\sqrt{2}\pi}\, \phi_+(x=0), \cr
\mathcal{H}_1 + \mathcal{H}_2 & = \mathcal{H}_{0+} + \mathcal{H}_{0-},
\end{align}
where the Hamiltonians $\mathcal{H}_{0 \pm}$ (in terms of $\phi_{\pm}$) are given by
\begin{align}
\label{hamil:strong}
&\mathcal{H}_{0 \pm} = \frac{(v_F + g_4/2\pi)}{4\pi}\,\int_{-L/2}^{L/2} dx 
(\partial_x \phi_{\pm})^2 \cr
&+\frac{g_2}{2(2\pi)^2}\,\int_{-L/2}^{L/2} dx dy \delta(x+y) \partial_x \phi_{\pm}(x) \partial_y \phi_{\pm}(y).
\end{align}
As can be seen in Eq.(\ref{allofthem}), the Zeeman coupling Hamiltonian $\mathcal{H}_{B,s}$ ( which is solely expressed 
in terms of $\phi_+$ )
is \textit{decoupled} from the 
tunneling Hamiltonian (which is solely expressed in terms of $\phi_-$), and this implies that 
\textit{the spin transport is not affected by the tunneling}.

The Hamiltonian Eq.(\ref{hamil:strong}) can be diagonalized by the following Bogoliubov transformation.\cite{bosobook}
\be
\label{transformation}
\phi_a(t,x) =  \varphi_a(t,x) \cosh \zeta - \varphi_a(t,-x)  \sinh \zeta .
\ee
After the diagonalization, the corresponding action for the chiral boson $\varphi_\pm$  is given by\cite{wen}
\begin{align}
\label{action:strong2}
S_{0\pm} &= - \frac{1}{4\pi}\,\int_{-\infty}^\infty dt \int_{-L/2}^{L/2} dx \partial_t \varphi_\pm
\partial_x \varphi_\pm \cr
&-\frac{v_0}{4\pi}\,\int_{-\infty}^\infty dt \int_{-L/2}^{L/2} dx 
(\partial_x \varphi_{\pm})^2,
\end{align}
where $K$ and $v_0$ are the same LL parameter and the velocity of collective excitation in the weak scattering regime
given in Eq.(\ref{parameters}). The Bogoliubov parameters are
\be
\label{cosh}
\cosh \zeta = \frac{K+K^{-1}}{2}, \quad \sinh \zeta = \frac{K-K^{-1}}{2}.
\ee
From Eq.(\ref{transformation},\ref{cosh}) we find that when the field is near the boundary 
\be
\phi_a(x \to 0, t) = \frac{1}{K} \varphi_a(x \to 0, t).
\ee
\underline{The spin transport}-
For the spin transport we can still apply the Keldysh formalism.
The spin current can be calculated by
\begin{align}
&J_\sigma(t) =i \mu_B^{-1} \frac{d}{dt}\, \la   e^{ i S_{B,s}}  \ra \Big \vert_{B_{0q} \to 0}, \cr
& S_B  =\frac{ \mu_B \kappa}{\sqrt{2}\pi} \int dt  \Big( B_{0q}(t) \phi_{c+}(t) + B_{0c}(t) \phi_{q+}(t) \Big ).
\end{align}
The calculation is entirely identical with that of the weak scattering case
\be
\label{main3}
J_\sigma(t) =  \frac{ \kappa^2 \mu_B}{2\pi K} B_0,\;\;\text{no corrections},
\ee
where we have used $\la \phi_{c+}(t_1) \phi_{q+}(t_2) \ra = - i \pi K \Theta(t_1 - t_2)/2$.
This result is the \textit{same as that of the weak scattering regime}.
It indicates that the impurity is basically decoupled from the spin degrees of freedom.
The spin conductance is 
\be
\label{main3-results}
G_\sigma = \frac{J_\sigma}{B_0} \Big \vert_{B_0 \to 0} = \frac{ \mu_B \kappa^2}{2\pi K},\;\;
\text{no corrections},
\ee
We can ask how the finite spin conductance is possible at zero temperature. At $T = 0$ fixed point,
 basically all the right movers are reflected into the left movers. However, as mentioned previously, 
the spin does not see the boundary since the orientation of the spin remains the same either in the presence or 
in the absence of the boundary.

\underline{The charge transport}- the charge current is given by
\be
J_\rho(t) = e \frac{ d \la \widehat{N}_- (t) \ra}{d t}.
\ee
The time-dependence of $ \widehat{N}_-$ solely comes from the tunneling Hamitonian 
$H_T$. An efficient way of treating the dynamics of  $ \widehat{N}_- $ and the zero modes 
is discussed in Ref.[\onlinecite{balents}].
It is clear the only non-vanishing contribution to current comes from the second order in tunneling
Hamiltonian. In the interaction picture with respect to $\mathcal{H}_- + \mathcal{H}_{V,s}$,
\be
(\mathcal{H}_{T})_{I}(t) = t_u \Big[ e^{ -i e V t} F e^{i \sqrt{2} \phi_-(0,t)} +\mathrm{H.c} \Big ].
\ee
The time evolution of $\phi_-(0,t)$ is implicitly assumed and the subscript $I$ is omitted.
Since the Klein factor and the number operator $\widehat{N}_-$ do not obey the canonical commuation relation,
the direct application of Keldysh path integral is not feasible. Instead, it is better to evaluate the
expectation values directly in the Dyson expansion of time dependent perturbation theory.
A straightforward calculation, employing $F^\dag \widehat{N}_- F = ( \widehat{N}_- -1)$ and $F \widehat{N}_- F^\dag = ( \widehat{N}_- +1)$, shows that
\begin{align}
 \la \widehat{N}_-(t) \ra & = 2 t_u^2 \int_{-\infty}^t d t_1 \int_{-\infty}^t d t_2\,
e^{-2 C(t_1-t_2)/K} \cr
&\times \sin [eV(t_1-t_2)] \sin [ \frac{\pi}{K} \mathrm{sign}(t_1-t_2) ],
\end{align}
where  $C(t)$ is given by Eq.(\ref{C-function}).
Now using the explicit result of $C(t)$  we get
\be
\label{main4}
J_\rho(t) \sim  \, e t_u^2 \int_0^\infty d t' \,\frac{ \sin (e V t')}{
(t^{\prime 2}+\tau_c^2)^{1/K} ( \frac{ \sinh ( t' \pi /\beta)}{t' \pi /\beta} )^{2/K}}.
\ee
From the above result the charge conductance at finite $T$ easily follows:
\be
\label{main4-results}
G_\rho = c_1' e^2 t_u^2 T^{2/K-2},
\ee
where $c_1'$ is a constant.
Now let us our compare results Eq.(\ref{main3-results},\ref{main4-results}) with those of  the spinful LL.
The charge and the spin conductance of spinful LL in strong scattering regime are given by (see Eq.(4.21) and Eq.(4.26) of 
Ref.[\onlinecite{akira}])
\begin{align}
\label{LL:strong}
G_\rho(T) &\sim d_1 e^2 t^2 T^{\frac{1}{K_\rho}+\frac{1}{K_\sigma}-2},\cr
G_\sigma(T) &\sim d_2 \mu_B t^2 T^{\frac{1}{K_\rho}+\frac{1}{K_\sigma}-2},
\end{align}
where $d_{1,2}$ are constants.
Again the charge transport of 1D ASF in the strong scattering regime is consistent with that of \textit{spinless} LL, while
the spin transport is radically different. 
The spin current of the spinful LL is degraded by impurities while that of ASF is not.
\section{Summary and Discussions}
In this paper, we have investigated the  effects of an spinless 
impurity on the transport properties of 1D ASF. 
Due to the strong spin-charge mixing effect, the spin transport is not affected by the impurity
, which is radically different from that of ordinary spinful LL where the spin current is equally strongly degraded 
by impurity as the charge current. On the other hand, the charge 
transport is essentially  identical with  that of the spinless LL.
The results of this paper can be verified by direct transport measurements, or by the 
recently developed momentum-selective tunneling transport measurements.\cite{governale2002,boese2001,auslaender2002}

\begin{acknowledgements}
This work was supported  by the Grant No. R01-2005-000-10352-0 from the Basic Research Program of the Korean Science and Engineering Foundation and by the Korea Research Foundation Grant funded by the Korean Government (MOEHRD)(KRF-2005-070-C0044).
\end{acknowledgements}



\begin{thebibliography}{}
\bibitem{note0} Note that in Eq.(\ref{operator}) the subleading corrections proportional to $|k\pm k_F|$ arising from the 
linearization of matrix elements $u_k^-$and $v_k^-$ are ignored at low energy.
\bibitem{boundaryterm} 
All of the oscillating terms of the type $e^{\pm i 2 k_F x}$  are neglected upon spatial integration. Also,  the boundary terms $\theta(\pm L/2)$ are irrelevant to  transports and are set
to zero.
\end{thebibliography}
\end{document}